# Effective electro-optic modulation in low-loss graphene-plasmonic slot waveguides

Yunhong Ding*[1], Xiaowei Guan[1], Xiaolong Zhu[3], Hao Hu[1], Sergey I. Bozhevolnyi[4], Leif Katsuo Oxenløwe[1], K. J. Jin[5], N. Asger Mortensen[1,2,4], and Sanshui Xiao†[1,2]

(1) Department of Photonics Engineering, Technical University of Denmark, DK-2800 Kongens Lyngby, Denmark
(2) Center for Nanostructured Graphene, Technical University of Denmark, DK-2800 Kongens Lyngby, Denmark
(3) Department of Micro- and Nanotechnology, Technical University of Denmark, DK-2800 Kongens Lyngby, Denmark
(4) Centre for Nano Optics & Danish Institute for Advanced Study, University of Southern Denmark, Campusvej 55, DK-5230 Odense M, Denmark
(5) Beijing National Laboratory for Condensed Matter Physics, Institute of Physics, Chinese Academy of Sciences, Beijing, 100190, China

*yudin@fotonik.dtu.dk, †saxi@fotonik.dtu.dk

**Surface plasmon polaritons enable light concentration within subwavelength regions, opening thereby new avenues for miniaturizing the device and strengthening light-matter interactions. Here we realize effective electro-optic modulation in low-loss plasmonic waveguides with the aid of graphene, and the devices are fully integrated in the silicon-on-insulator platform. By advantageously exploiting low-loss plasmonic slot-waveguide modes, which weakly leak into a substrate while feature strong fields within the two-layer-graphene covered slots in metal, we successfully achieve a tunability of 0.13 dB/µm for our fabricated graphene-plasmonic waveguide devices with extremely low insertion loss, which outperforms previously reported graphene-plasmonic devices. Our results highlight the potential of graphene plasmonic leaky-mode hybrid waveguides to realized active ultra-compact devices for optoelectronic applications.**

Due to its unique electronic and optical properties, graphene has offered a new paradigm for extremely fast and active optoelectronic devices [1-7], e.g., the tunability of graphene conductivity [8-10] allows to realize efficient electro-optical (E/O) modulation [11], and all-optical switching [12,13]. With the combination of high-index dielectric waveguides/resonators, several integrated graphene-based optical modulators have already been realized [14-21]. However, the optical modes in these systems are inherently strongly localized in the high-index materials, thus jeopardizing light-graphene interactions. Moreover, the size of the waveguide in these systems is bounded by the diffraction limit, thus hindering the miniaturization of the devices. Surface plasmon polaritons (SPPs) are broadband with the ability to manipulate light on the subwavelength scale [22-25], while at the same time giving possibility to direct more optical energy to the material interface where graphene could reside. The strong confinement of SPP allows realizing compact optical devices with low energy consumption, and enables more optical energy to enter graphene-light interactions. Hybrid graphene plasmonic waveguides have been explored for E/O modulation [26,27], while less transmission loss and more significant modulation depth are required for optical communications. Although many applications benefit from the subwavelength confinement of SPPs, the inevitable sensitivity to imperfections puts strict fabrication demands on conventional SPP designs and there is a strive for SPP that can travel over large distances, while preserving strong light-matter interaction [28].

Leaky modes refer to quasi-confined modes that are losing energy when propagating, and they are usually prevented in conventional dielectric photonics. In a plasmonic waveguide system, leaky modes are intimately linked to radiation loss [29, 30]. In the meantime, leaky modes also push the optical field away from the metal, leading to lower Ohmic loss, especially for a metal-slot plasmonic waveguide where the optical fields extremely interact with the relatively rough slot sidewalls. Consequently, the total loss of the leaky-mode plasmonic waveguide is reduced. This phenomenon is quite different with the conventional dielectric waveguide where the leaky mode can only exacerbate the loss. Here, we for the first time propose and demonstrate efficient E/O modulating based on leaky-mode plasmonic slot waveguides, which are fully integrated with the silicon-on-insulator (SOI) platform. Both graphene tunability and propagation loss are similarly affected by the mode confinement, but the relevant regions of confinement are different. This opens a way to optimizing the trade-off between the propagation loss and the tunability by propitious design that leads us to exploiting a leaky-mode regime of the slot waveguide. We experimentally demonstrate low loss of the plasmonic slot waveguide, and achieve high tunability for the integrated graphene based plasmonic E/O devices covered with the two-layer graphene.

**Results**

**Design of graphene plasmonic waveguide devices.**

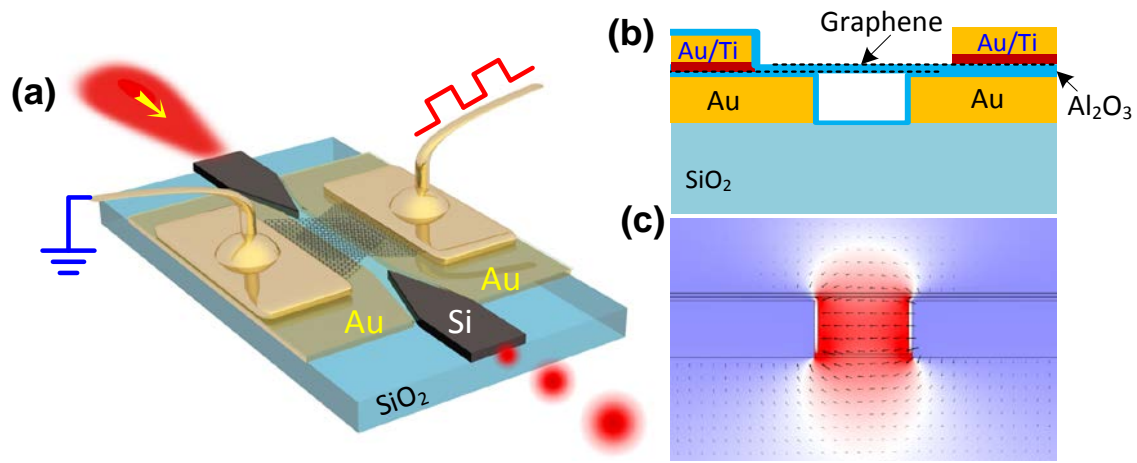

**Figure 1. Modulator structure and waveguide mode profile**. (a) 3D schematic of the designed graphene-plasmonic slot hybrid waveguide based E/O modulator. (b) The cross-section of the waveguide, and (c) the corresponding plasmonic slot mode-field profile, indicating how field intensity is being strongly confined to the slot region.

The effective E/O modulation of a graphene-plasmonic slot waveguide is based on tuning of the Fermi level of two-layer graphene on top of the waveguide. Figure 1(A) shows three-dimensional schematic of the proposed device. The plasmonic slot waveguide is formed by introducing a small slot in a thin gold (Au) sheet, and coupled with silicon waveguides with inverse tapering tips [31]. Two layers of graphene are overlapped and placed on top of the plasmonic slot waveguide, and a thin layer of Al2O3 is introduced between them. The graphene- Al2O3-graphene sandwich forms an effective capacitor. When the capacitor is biased, carriers can be accumulated on one graphene sheet and swept out from another, resulting in an efficient tuning of the Fermi level of both graphene layers. In turn, the deliberately doped graphene layers now allow to tune optical absorption in the plasmonic waveguide. An extra thin layer of Al2O3 is introduced between the lower graphene sheet and the Au plasmonic waveguide to electrically isolate the contact between them. The gold/titanium (Au/Ti) contacts are used for electrical contacts with the two-layer graphene sheets, and are designed to be 3 μm away from the central plasmonic slot waveguide to suppress undesired absorption loss associated with the metals. Figure 1(B) presents the cross-section for one of the designed waveguides (with air being the upper cladding) and the characteristic transverse electric (TE) mode profile at 1.55 μm of the plasmonic slot is shown in Fig. 1(C), calculated with the aid of a commercial available package (COMSOL RF Module). All parameters used in the simulation can be found in the Supplementary Materials. The field is strongly confined to the slot region, thus paving the way for strong graphene-light interactions.

**Analysis of light-graphene interaction in graphene plasmonic hybrid waveguide.**

Regarding the plasmonic slot waveguide design, see the Supporting Materials, here we focus on three different generic cross-section of plasmonic slot waveguides: i.e. semi-symmetrical arrangement with SiO2 as the low cladding and PMMA as the upper cladding, symmetrical waveguide with air as both upper and lower claddings, and asymmetric waveguide with air being the upper cladding and SiO2 being the low cladding. The former two types of plasmonic slot waveguides often support guided modes, while the latter air-SiO2 asymmetric waveguide supports leaky modes above the cutoff slot width or film thickness [29, 32]. For reazling effective E/O modulating, it is crucial to have a large modulation depth, while keeping the insertion loss as low as possible. The trade-off between the propagation loss and tunability in these waveguide configurations has to be taken into account. Here we systematically analyze the light-graphene interactions in these systems with respect to tunability and modulation depth. Figure 2(A) shows the transmission as a function of the Fermi level for the three kinds of waveguides interacting with two sheets of graphene. The details of these two-layer graphene based plasmonic waveguides and the calculation models are presented in the Supporting Materials. The tunability (with respect to the Fermi level tuned from 0 to 0.4 eV) of 0.069 dB/μm, 0.18 dB/μm and 0.24 dB/μm are obtained for the air-SiO2, SiO2-PMMA, and air-air waveguides, respectively. The largest tunability is obtained for the case of the air-air waveguide, however at the price of a relative large insertion loss of 0.7 dB/μm. On the other hand, the air-SiO2 asymmetric structure has a relatively lower tunability due to light field distribution in the SiO2 cladding, while in return offering a more favorable loss of 0.1 dB/μm. When considering the same insertion loss of -10dB, the air-SiO2 asymmetric waveguide surprisingly gives the largest modulation depth, see Fig. 2(B). Here we consider the slot gap of 120 nm and the Au thickness of 90 nm for all three waveguides at 1.55 μm.

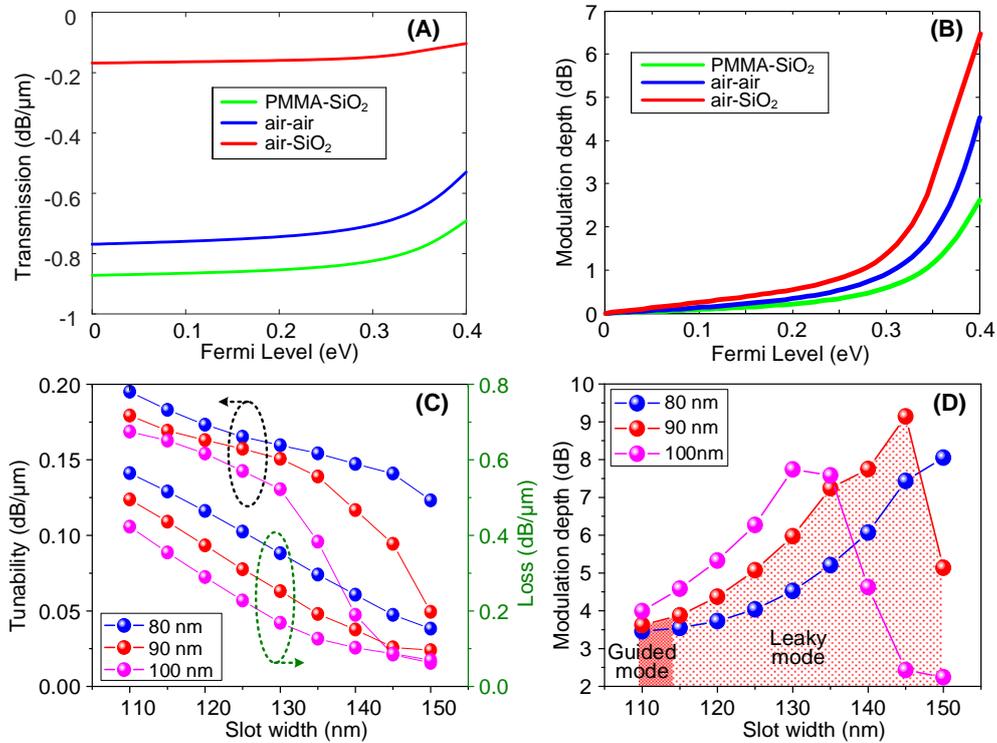

**Figure 2. Simulated results for graphene plasmonic hybrid waveguides.** (A) Transmission as a function of Fermi level for the three kinds of graphene-plasmonic slot waveguides. (B) Corresponding modulation depths when considering the same insertion loss of -10 dB with the slot gap of 120 nm and Au thickness of 90 nm. (C) Tunability (with respect to the Fermi level tuned from 0 to 0.4 eV) and loss of the asymmetric air-SiO2 plasmonic waveguide as a function of slot width for different thickness (80 nm, 90 nm, 100 nm) of Au film. (D) Structural parameter dependence of the modulation depth when considering the insertion loss of 10 dB, showing the optimum structure with respect to the highest modulation depth.

The leaky feature in the asymmetric waveguide weakens the light-graphene interaction, resulting in low tunability. However, it also enables to distribute optical field away from the metal, thus decreasing the Ohmic loss. Balancing these two factors, the asymmetric air-SiO2 waveguide design eventually offers the largest modulation depth when having the otherwise same insertion loss. Moreover, this waveguide configuration is also the simplest from the fabrication point of view. For this asymmetric waveguide configuration, the tunability and loss as a function of slot size are investigated and results are summarized in Fig. 2(C) for different Au flim thickness with 80 nm, 90 nm, and 100 nm, respectively.

The tunability decreases as the slot width increases. In addition, a thicker Au film also results in a lower tunability. This trend is attributed to the fact that the field redistributes further away from the graphene sheets when either the slot width or the Au thickness increases. On the other hand, the loss of the plasmonic slot waveguides also decreases with the increase of slot width and Au film thickness. When considering the insertion loss of 10 dB, Fig. 2(D) shows that for each film thickness there is an optimum slot width that provides the highest modulation depth (defined by 10/Loss*Tunability). As the Au thickness increases, the optimum size of the slot width decreases. For the Au film thickness of 90 nm, a highest modulation depth of 9 dB can be obtained with the slot width of 145 nm. If the gap size is below 114 nm for the case of 90 nm-thick Au film, the waveguide supported a guided mode, while the modulation depth is much smaller than that in the leaky-mode regime.

**Graphene electro-optical modulator based on plasmonic leaky-mode-slot waveguide.**

The device was fabricated on a commercially available SOI wafer with a top-silicon layer of 250 nm and a buried oxide layer of 3 μm. The top-silicon layer was p-type with a resistivity of ~20 Ω·cm. Firstly, the top silicon layer was thinned down to 90 nm. Then standard SOI processing, including e-beam lithography (EBL) and inductively coupled plasma etching was used to fabricate the silicon waveguides. The 12 μm-wide apodized grating couplers [31] with a 400 μm long adiabatic taper were used to couple between the silicon waveguides and standard single mode fibers. Subsequently, the plasmonic slot waveguide was defined by the second EBL, and followed by Au deposition and liftoff processing. After that, a ~5 nm thin Al2O3 layer was grown by atom-layer deposition (ALD) machine on the chip to isolate the Au plasmonic slot waveguide and the graphene sheet that will be transferred later. Then, a 1.5 cm×1.5 cm graphene sheet (grown by CVD) was wet-transferred [33, 34] on top of the chip. In this step, a photoresist AZ5214E, was firstly spin-coated onto the graphene covered copper foil and then dried at 90 °C for 1 min. Subsequently, a AZ5214/graphene membrane was obtained by etching away the copper foil in a Fe(NO3)3/H2O solution and then transferred onto the silicon chip. Finally, the AZ5214E was dissolved in acetone that also served to clean the graphene surface. The graphene-covered areas on the plasmonic slot waveguides were then defined by standard ultraviolet (UV) lithography followed by oxygen (O2) plasma etching. Following that, the contact windows on graphene were defined by UV lithography, and Au/Ti contacts on the graphene were then obtained by Au/Ti metal deposition and a liftoff process. Afterwards, 1 nm Al was deposited on the chip, and oxidized to Al2O3 in air. This native Al2O3 layer works as a seed layer for the further Al2O3 deposition in the ALD machine. Eventually, 10 nm Al2O3 was obtained on top of the first graphene layer. The second graphene layer was wet-transferred with the same transferring process. The coverage areas of the second graphene layer were formed by standard UV lithography followed by oxygen plasma etching. Finally, the Au/Ti contacts on the second graphene layer were fabricated by UV lithography followed by Au/Ti metal deposition and a liftoff process.

Following the optimization of plasmonic leaky-mode slot wave-guides, we fabricate and characterize graphene-plasmonic hybrid waveguides, see Supporting Information Sections 4 and 5. Photonic crystal grating couplers [35] aid coupling light in and out of the devices. Figure 3(A) exhibits the fabricated device with 4 μm-long Au plasmonic slot waveguide, which is coupled with 600 nm-wide and 90 nm-height silicon waveguides with inverse tapering tips. The zoom-in, see Fig. 3(B), of the coupling part from the silicon waveguide to the plasmonic slot waveguide shows good alignment between the silicon and plasmonic waveguide, leading to high in/out coupling loss of 1.45 dB. This proposed coupling platform has a very good alignment tolerance (see Supporting Materials). High quality of the transferred two-layer graphene is clearly shown as well. The measured transmission spectra for the asymmetric plasmonic slot waveguides without graphene are presented in Fig. 3(C) as a function of waveguide length, showing broadband feature of the plasmonic waveguide. Here the slot width is 120 nm, and Au film thickness is 90 nm, respectively. A linear propagation loss of 0.25 dB/µm, see the blue line in Fig. 3(D), is obtained by the cut-back method, which improves over guided-mode plasmonic slot waveguides [36, 37]. The transmission of –2.9 dB when L→0 µm shown in Fig. 3(d) indicates a coupling loss of 1.45 dB between the silicon and plasmonic slot waveguides, which agrees well with our theoretical prediction (see Supporting Materials). The linear loss of the final fabricated plasmonic slot waveguide with the two-layer graphene is relatively high, 0.68 dB/µm shown by the red line in Fig. 3(D). This value is much higher than that anticipated by simulations (0.18 dB/µm, see Supporting Materials), and we believe that excess loss is arisen from the residual contaminations during the wet transferring process and the surface roughness for the plasmonic slot waveguide. More clean wet-transferring processing, e.g. modified RCA process associated wet-transferring and optimization of nanofabrication, would further decrease the loss.

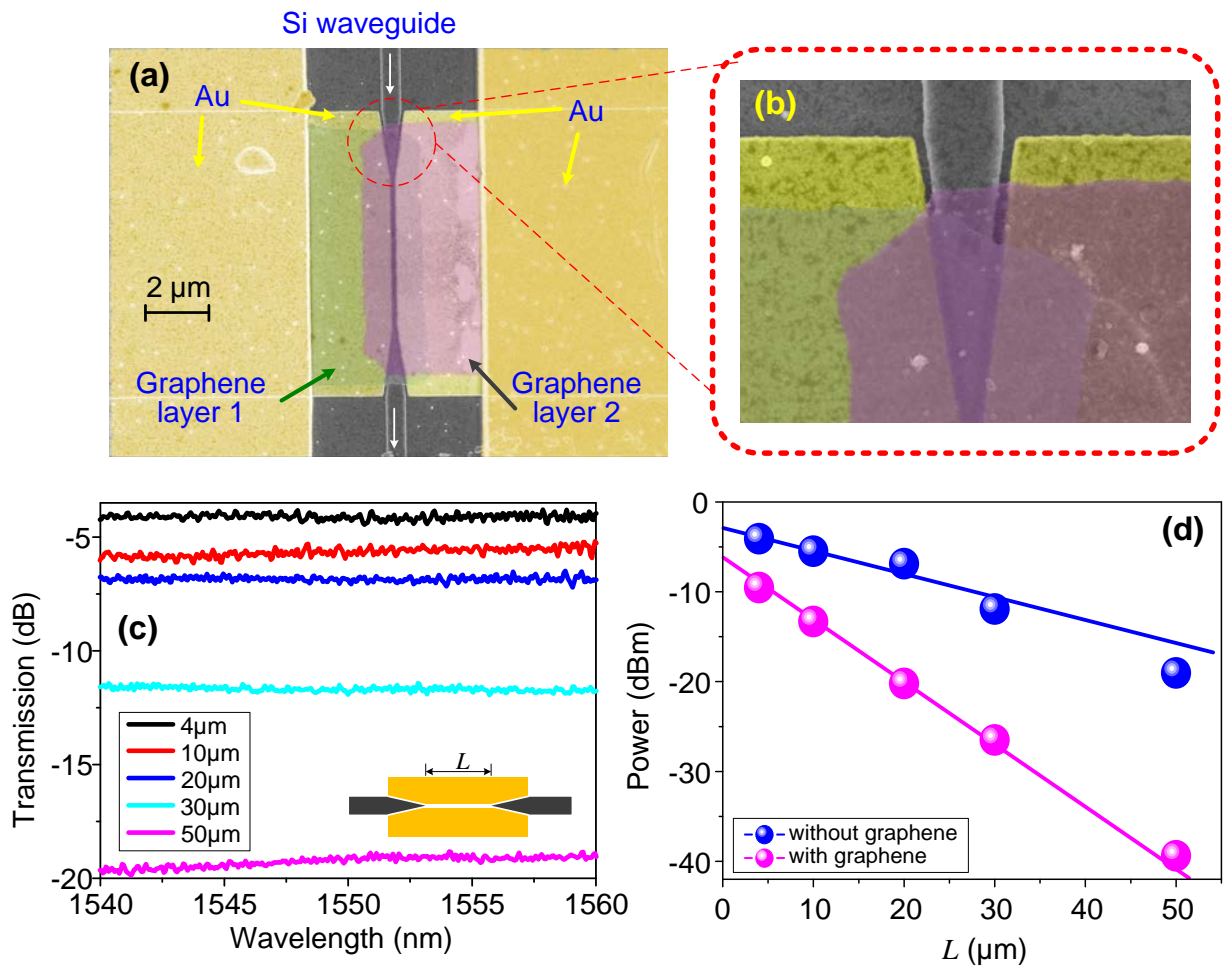

**Figure 3. Graphene plasmonic waveguide modulator and plasmonic slot waveguide** (a) False-color SEM image of the fabricated graphene-plasmonic slot hybrid waveguide. (b) A zoom-in on the coupling part of the fabricated device, clearly indicating good alignment between the silicon and plasmonic waveguides and good quality of the transferred two-layer graphene sheet. (c) Measured transmitted optical power of the plasmonic slot waveguide without graphene as a function of wavelength for different slot lengths $L$. (d) Cut-back measurements of the plasmonic slot waveguide without and with graphene at wavelength of 1.55 µm. The slope of the blue line indicates a loss of 0.25 dB/µm and the coupling loss of 1.45 dB for the plasmonic waveguide without graphene, and the slope of the purple line indicates a loss of 0.68 dB/µm for the graphene-plasmonic hybrid waveguide.

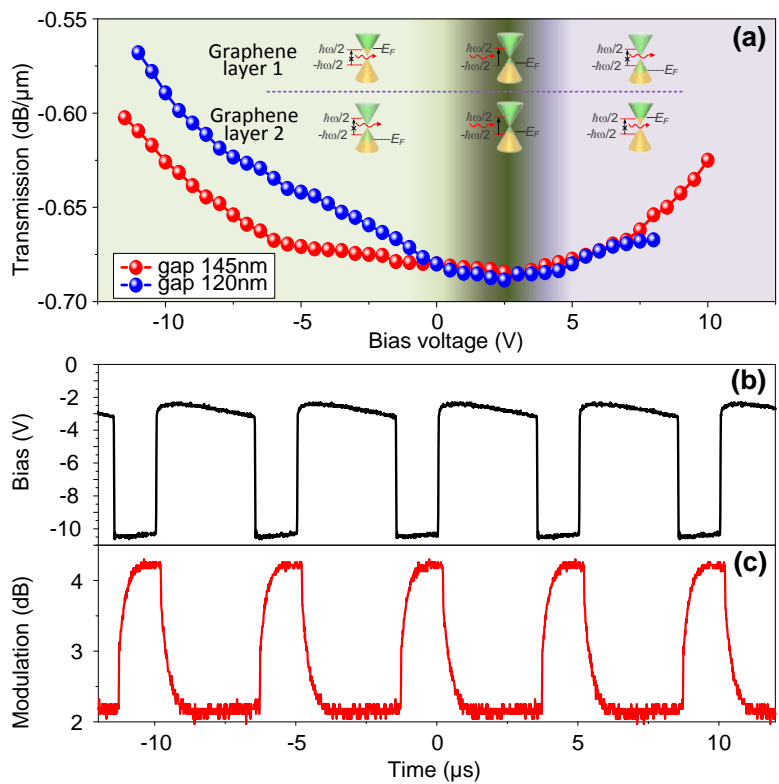

**Figure 4. Performance of graphene plasmonic waveguide modulator.** (a) Measured transmission loss of the graphene-silicon waveguide as a function of bias voltage for 20 µm-long plasmonic slot waveguide with plasmonic gap of 145 nm and 120 nm. (b) Applied electrical square waveform with a peak-to-peak voltage of 7.5 V with bias voltage of -6.75 V. (c) Measured modulated optical signal with an on-off ratio of 2.1 dB.

Figure 4(A) displays the light (1.55 µm) transmission through 20 µm-long leaky-mode graphene-plasmonic waveguides at different bias voltages for two slot widths of 120 nm and 145 nm, respectively. We find that the transmission of light through the graphene-plasmonic hybrid waveguides is effectively tuned by applying bias voltages on the two-layer graphene. When the biasvoltage is low (-0.5 V to 4 V), the Fermi levels EF of two graphene layers are close to the Dirac point (EF<$\hbar\omega/2$), resulting in high damping due to inter-band transitions in the graphene for the photons (with energy $\hbar\omega$) propagating throughthe waveguide. When a higher negative voltage (<-0.5 V) is applied, the Fermi level of the graphene layer 1 is lifted (>$\hbar\omega/2$) while the Fermi level of the graphene layer 2 is lowered (<$-\hbar\omega/2$). Consequently, light propagation is less attenuated since no electrons are available for inter-band transitions for the graphene layer 2, at the same time all electron states are filled up in the graphene layer 1 and no inter-band transition is allowed due to Pauli blocking. On the other hand, a high positive voltage (>4 V) results in a lower Fermi level (<$-\hbar\omega/2$) for the graphene layer 1 and higher Fermi level (>$\hbar\omega/2$) for the graphene layer 2. In this case, a lower optical propagation loss is also obtained since no electrons available in the graphene layer 1 and all electron states are filled up in the graphene layer 2. The dip position at 2.5 V in the transmission spectra shows the charge neutrality point in graphene, i.e. the CVD-graphene is initially doped. An efficient attenuation tunability of 0.13 dB/µm (from 0.69 dB/µm to 0.56 dB/µm) is achieved for the plasmonic slot width of 120 nm at low gating voltages. When the graphene is negatively biased, smaller tunability of 0.08 dB/µm (from 0.68 dB/µm to 0.6 dB/µm) is obtained for the plasmonic slot width of 145 nm, which is well consistent with the theoretical analysis, i.e., the graphene-light interaction becomes weaker for a wider plasmonic slot width. The larger modulation depth of 0.13 dB/µm achieved in our leaky-mode plasmonic waveguide device incorporated with two-layer graphene is higher than previously realized graphene-plasmonic devices [26]. We emphasize that for the graphene plasmonic waveguide device, the modulation depth can be improved by use of a guided-mode plasmonic waveguide, however it will suffer significantly by large insertion loss. The graphene-plasmonic based E/O device with the slot width of 120 nm is further tested by applying a square waveform bias in order to explore the dynamic switching at

1.55 µm, see Fig. 4(B). The square waveform with a peak-to-peak voltage of 7.5 V and bias of −6.75 V results in an effective optical switch with an extinction ratio of 2.1 dB, as shown in Fig. 4(C). Higher extinction ratio can be obtained with larger peak-to-peak voltage for the driving signal.

Leaky plasmonic modes are commonly disregarded, but they offer new opportunities in combination with graphene for E/O modulating. Relying on graphene-plasmonic leaky-slot-mode waveguides, we have proposed and demonstrated graphene plasmonic waveguide for effective E/O modulating fully integrated with the SOI platform. We have comprehensively analyzed the interaction between graphene and the plasmonic slot waveguides, showing that the leaky-mode based waveguide provides much better modulation depth when considering the same insertion loss. We have experimentally achieved a considerable low loss of 0.68 dB/µm for the graphene-based plasmonic waveguide, and the tunability of 0.13 dB/µm for the graphene plasmonic modulator with low insertion loss. The result presented here provides a promising way to promote the plasmonic leaky-mode waveguide for improving the performance of graphene-plasmonic based optoelectronic devices. Our current efforts have focused on the modulation depth, while the operation speed can be optimized by dedicated fabrication efforts to reduce the RC time constant of the circuit. Since the optical waveguide mode is well concentrated within the gap size of ~145 nm, the resistance can be significantly reduced by designing the contact to be close to the slot waveguide. On the other hand, the capacitance can be reduced by decreasing the overlap area of the two-layer graphene sheets or increasing the thickness of the dielectric material Al2O3.

**Acknowledgment**

This work is supported by the Danish Council for Independent Research (DFF-1337-00152 and DFF-1335-00771). The Center for Nanostructured Graphene is sponsored by the Danish National Research Foundation, Project DNRF103. This work is also partly supported by the CAS/SAFEA International Partnership Program for Creative Research Teams (Project no. 20140491513) and the International Network program (Project no. 6144-00098). N.A.M is a VILLUM Investigator supported by VILLUM FONDEN (grant No. 16498).